# Balancing Innovation and Ethics in AI-Driven Software Development

Mohammad Baqar (mobaqar@cisco.com)

Security Business Group, Cisco Systems Inc, San Jose, CA, US

## Abstract

This paper critically examines the ethical implications of integrating AI tools like GitHub Copilot and ChatGPT into the software development process. It explores issues such as code ownership, bias, accountability, privacy, and the potential impact on the job market. While these AI tools offer significant benefits in terms of productivity and efficiency, they also introduce complex ethical challenges. The paper argues that addressing these challenges is essential to ensuring that AI's integration into software development is both responsible and beneficial to society.

# 1.  Introduction

## 1.1 Introduction to AI Tools in Software Development

AI tools like **ChatGPT** and **GitHub** Copilot represent a significant advancement in the software development process, offering developers powerful new ways to write, debug, and optimize code. These tools leverage large language models (LLMs) trained on vast amounts of data, including code repositories, technical documentation, and natural language text, to assist developers in real-time.

**GitHub Copilot,** developed by GitHub in collaboration with OpenAI, is an AI-powered code completion tool that functions as a "pair programmer." It is built on the Codex model, a descendant of GPT-3, specifically fine-tuned for programming tasks. GitHub Copilot assists developers by:

- Autocompleting code: As developers write code, GitHub Copilot suggests entire lines or blocks of code, predicting what the developer intends to write next based on the context.
- Generating boilerplate code: Copilot can generate repetitive or boilerplate code, saving developers time and reducing the risk of errors.
- Suggesting alternative implementations: It can propose different approaches to solving a problem, offering developers options to choose from.

**Integration in Development:** GitHub Copilot is integrated directly into popular code editors like Visual Studio Code. It operates as an extension that interacts with the code in real-time, providing suggestions as the developer types. This integration makes GitHub Copilot a natural extension of the developer's workflow, offering continuous support without requiring them to switch contexts or tools.

**ChatGPT** is a conversational AI model developed by OpenAI, based on the GPT (Generative Pre-trained Transformer) architecture. It is designed to understand and generate human-like text, making it highly adaptable for various tasks, including assisting in software development. In the context of coding, ChatGPT can:

- Generate code snippets: Developers can describe the functionality they need, and ChatGPT can generate relevant code snippets in various programming languages.
- Provide explanations and documentation: ChatGPT can help explain complex code, provide documentation, and answer technical questions, aiding both novice and experienced developers.
- Assist in debugging: By analyzing code, ChatGPT can suggest potential fixes for bugs and optimization strategies, helping developers troubleshoot issues more efficiently.

**Integration in Development:** ChatGPT is typically integrated into development environments through APIs or as a plugin. It can be used in code editors, IDEs, or as part of a continuous integration/continuous deployment (CI/CD) pipeline to provide real-time support. This integration allows developers to interact with ChatGPT directly from their development environment, making it a seamless part of the coding workflow.

## 1.2 Impact on the Software Development Process

The integration of AI tools like ChatGPT and GitHub Copilot into software development has several profound impacts:

- Increased Productivity: By automating repetitive tasks and providing instant suggestions, these tools help developers work more efficiently, allowing them to focus on more complex and creative aspects of coding.
- Reduced Errors: AI tools assist in identifying potential issues and suggesting corrections, leading to cleaner, more reliable code.
- Enhanced Collaboration: These tools can act as a bridge between team members with different levels of expertise, providing consistent guidance and improving overall team performance.

# 2.  Ethical Implications on Code Ownership

## 2.1 Authorship and Intellectual Property in AI-Generated Code

As AI tools like GitHub Copilot and ChatGPT become increasingly prevalent in software development, questions about authorship and intellectual property (IP) ownership of AI-generated code have emerged as critical legal and ethical concerns.

Who Owns AI-Generated Code?

The question of ownership centers on whether the code generated by AI is considered the intellectual property of the developer who used the tool, the organization employing the developer, or the creators of the AI tool itself. Traditionally, the creator of a work is recognized as its owner, but AI blurs these lines because the code is often the result of collaboration between human input and machine generation.

## The Developer's Claim:

- Developers using AI tools might argue that they retain ownership because they provide the initial input, guidance, and decision-making necessary to produce the final code. The AI serves as an assistant, similar to an advanced version of an IDE or text editor, rather than an independent creator.

- Developers typically refine, edit, and adapt AI-generated suggestions, which could further strengthen their claim to ownership. They might view the AI as a tool that facilitates the creative process, with the ultimate responsibility and ownership resting with them.

### The Organization's Perspective:

- In many employment contracts, particularly in the tech industry, the organization holds the IP rights to any work produced by its employees during the course of their employment. This often extends to code written using company-provided tools, including AI-driven ones.
- Organizations might assert that since they provide the resources (including access to AI tools), the code generated belongs to them. This aligns with the notion that the AI is a productivity enhancer, much like any other software or tool provided by the employer.

### The AI Tool Creators:

- There is also a potential argument from the creators of the AI tools, such as OpenAI or GitHub, claiming partial ownership or licensing rights to code generated by their models. This argument could be based on the proprietary nature of the AI's underlying algorithms and the extensive datasets used to train the models.
- However, this position is complex and controversial, as it challenges traditional notions of ownership and could discourage the widespread adoption of AI tools if developers fear losing their IP rights.

## Legal Precedents and Future Frameworks

As of now, the legal landscape regarding AI-generated works remains underdeveloped. Most jurisdictions do not have specific laws addressing the ownership of AI-generated code, leading to ambiguity and potential disputes.

### Current Legal Ambiguity:

- In most legal systems, IP law is based on the concept of human authorship. AI-generated content, therefore, falls into a gray area, as the AI itself cannot be considered a legal entity with ownership rights.
- There are precedents in other creative fields, such as art and music, where AI-generated works have led to debates about authorship. However, these cases have not yet resulted in clear, universally accepted legal standards.

### Possible Future Legal Frameworks:

- Human-Machine Collaboration Model: One potential legal framework could recognize AI-generated code as a collaborative effort between the developer and the AI tool, with shared or joint ownership. This could lead to new licensing models where developers and AI tool providers share rights or profits.
- Employer Ownership: Another approach might reinforce the employer's ownership of AI-generated code, especially if the code is produced within the scope of employment. This would extend current employment IP practices to include AI-assisted work.

- AI as a Service Provider: Alternatively, AI tools could be treated as service providers, with ownership of the output defaulting to the developer or organization using the tool, provided they comply with the terms of service or licensing agreements set by the AI tool providers.

## Implications for Software Development

The uncertainty surrounding IP rights in AI-generated code could have significant implications for software development. Developers may be hesitant to adopt AI tools if they are unclear about their rights to the code they produce. Organizations might need to reassess their IP policies and contracts to account for AI-assisted development, ensuring that ownership issues are clearly addressed.

Ultimately, the resolution of these ownership questions will likely require legal reform and possibly new international treaties or guidelines, as AI becomes an integral part of the software development landscape. Until then, developers, organizations, and AI providers must navigate this evolving area with caution, seeking to establish clear agreements and understandings about the ownership of AI-generated code.

## 2.2 Contribution vs. Generation in AI-Generated Code

The distinction between contribution and generation is a critical aspect of the debate surrounding AI-assisted software development. It addresses the varying levels of involvement and control that developers and AI tools have over the code produced. Understanding this distinction is key to determining ownership, responsibility, and ethical considerations in AI-generated code.

## Contribution: The Developer's Role

In many cases, AI tools like GitHub Copilot and ChatGPT act as assistants rather than autonomous creators. The developer's input, decision-making, and guidance play a crucial role in the final code output. This process can be described as a contribution, where the AI tool offers suggestions, which the developer then reviews, modifies, or rejects.

### Human Agency and Creativity:

- The developer remains the primary agent, using their knowledge and creativity to decide which AI-generated suggestions to implement. The AI provides potential solutions or improvements, but the developer makes the final call on what code to use.
- This scenario is akin to a brainstorming session where the AI serves as a knowledgeable assistant, offering ideas that the developer considers and adapts to fit the specific requirements of the project.

### Shared Ownership:

- In cases where the AI's contribution is significant but not wholly autonomous, there may be an argument for shared ownership of the code. The developer's active role in shaping and refining the AI's output suggests that the final product is a collaborative effort.
- However, even in these cases, many argue that the developer should retain primary ownership since they are the ones ultimately responsible for the integration and functionality of the code.

### Ethical Considerations:

- When the AI is contributing rather than generating code independently, ethical considerations focus on transparency and acknowledgment of the AI's role. Developers should be clear about which parts of the code were influenced by AI suggestions and ensure that these suggestions align with the project's ethical and functional standards.

## Generation: The AI's Role

In contrast, when AI tools generate entire blocks of code with minimal human intervention, the process shifts from contribution to generation. Here, the AI acts more autonomously, creating code that developers may use with little to no modification. This raises different questions about authorship, ownership, and responsibility.

### Autonomous Creation:

- AI tools like GitHub Copilot can generate complex code structures based on minimal input, such as a brief comment or a few lines of code. In these cases, the AI is not merely contributing to the development process but generating substantial portions of the code independently.
- The developer's role may be reduced to selecting and implementing the AI-generated code, potentially with only minor tweaks or no changes at all.

### Ownership and Intellectual Property:

- The ownership of code generated autonomously by AI is more contentious. If the AI is considered the primary creator, questions arise about whether the code can be owned by the developer, the organization, or even the AI tool provider.
- Some argue that if the AI is generating code without significant human input, the ownership could default to the AI tool provider, particularly if the generation process relies heavily on proprietary algorithms and datasets.

### Legal and Ethical Implications:

- Legal Uncertainty: The legal system currently lacks clear guidelines on the ownership of AI-generated works. This uncertainty could lead to disputes over who has the right to use, modify, or distribute AI-generated code.
- Ethical Responsibility: If an AI generates code that contains errors or leads to unintended consequences, determining who is ethically responsible becomes challenging. The developer, who implemented the code, or the AI provider, who created the tool, could both be held accountable, depending on the circumstances.

## Blurring the Lines: Hybrid Scenarios

In practice, the lines between contribution and generation are often blurred. AI tools might start by generating code, which developers then modify or integrate with their own contributions. This collaborative process raises further questions about how to attribute authorship and ownership.

### Hybrid Ownership Models:

- One possible approach is to recognize hybrid ownership, where both the developer and the AI tool provider have claims to the code, depending on the level of modification and integration by the developer.
- This model could lead to new forms of licensing agreements that acknowledge the dual contributions of human developers and AI tools.

### Ethical Transparency:

- In hybrid scenarios, transparency becomes crucial. Developers should disclose the extent to which AI tools were used in generating the code and whether significant modifications were made. This transparency is important for maintaining trust and ensuring that all contributors receive appropriate recognition.

The distinction between contribution and generation in AI-generated code has profound implications for ownership, intellectual property, and ethical responsibility. As AI tools continue to evolve, developers, organizations, and legal systems will need to adapt to these new realities, finding ways to fairly attribute ownership and responsibility while fostering innovation and creativity in software development.

## 3.  Bias in AI-Generated Code

## 3.1 Sources of Bias

AI tools like ChatGPT and GitHub Copilot rely on large datasets to learn patterns, generate text, and offer code suggestions. These datasets are often sourced from publicly available code repositories, online forums, documentation, and other forms of text that reflect the practices and language used by developers across the world. However, these datasets can contain inherent biases that the AI models may inadvertently learn and reproduce.

### Biased Training Data:

- The training data used for AI models might include code or documentation that reflects societal biases, such as gender stereotypes, racial prejudices, or economic inequalities. For example, if the training data includes code written by a predominantly male demographic, the AI might generate suggestions that align more closely with male-centric perspectives.
- Additionally, datasets can include outdated or discriminatory practices that are no longer considered acceptable, such as biased algorithms in criminal justice or biased language in documentation. The AI, trained on these datasets, might replicate these biases in its output.

### Imbalanced Representation:

- If certain groups are underrepresented in the training data, the AI might not learn to generate code that adequately addresses the needs of these groups. For instance, AI tools trained on datasets with limited representation of female developers might produce code suggestions that fail to account for diverse user needs or fail to recognize the contributions of underrepresented groups.

- This imbalance can lead to biased suggestions, such as assuming default roles, preferences, or behaviors that do not accurately reflect the diversity of potential users or developers.

## 3.2 Impact on Software Development

Biased AI-generated code can have significant consequences, particularly when used in sensitive applications where fairness, equity, and justice are paramount.

### Amplifying Societal Biases:

- When biased code is used in areas like criminal justice, healthcare, or finance, it can perpetuate or even amplify existing societal biases. For example, if AI-generated code is used in predictive policing algorithms, and the training data includes biased arrest records, the resulting code could reinforce racial profiling or other forms of discrimination.
- In healthcare, biased AI-generated code could lead to unequal treatment recommendations, disproportionately affecting minority groups. In finance, it could result in biased credit scoring algorithms, exacerbating economic disparities.

### Erosion of Trust:

- As AI tools become more integrated into software development, the presence of bias in AI-generated code can erode trust in these tools. Developers and end-users may become wary of relying on AI suggestions if they believe the tools are perpetuating harmful biases.
- This erosion of trust can slow the adoption of AI in critical industries and lead to a backlash against AI-driven innovations, undermining the potential benefits of these technologies.

## 3.3 Mitigation Strategies

To address bias in AI-generated code, developers, organizations, and AI providers must implement strategies that ensure fairness, equity, and transparency in AI models and their outputs.

### Diverse Training Datasets:

- One of the most effective ways to mitigate bias is to ensure that the training datasets used for AI models are diverse and representative of different populations, cultures, and perspectives. By including data from a wide range of sources, AI tools are more likely to generate code that is fair and unbiased.
- Organizations should actively seek out and include data from underrepresented groups, ensuring that the AI models learn from a variety of perspectives and avoid overfitting to the dominant majority.

### Algorithmic Transparency:

- Transparency in the algorithms and decision-making processes of AI tools is crucial for identifying and addressing bias. Developers should have access to information about how AI models make decisions and what data they were trained on, allowing for the identification of potential biases.
- AI providers can enhance transparency by providing detailed documentation, model interpretability tools, and regular audits to ensure that their algorithms do not produce biased outputs.

## Bias Detection and Correction:

- Implementing tools and processes to detect bias in AI-generated code is essential for ongoing monitoring and improvement. Techniques such as fairness-aware algorithms, adversarial testing, and bias detection metrics can help identify biased code suggestions.
- Once biases are detected, corrective measures such as re-training models on more diverse data, adjusting algorithms to counteract bias, and providing developers with options to override or modify biased suggestions can be employed.

## Ethical AI Development Practices:

- Adopting ethical guidelines and best practices for AI development is critical for reducing bias. This includes promoting diversity within development teams, encouraging ethical considerations during the design and implementation phases, and fostering a culture of responsibility and accountability in AI development.
- Organizations should also engage with external stakeholders, such as ethicists, civil society groups, and affected communities, to ensure that AI tools are developed and deployed in ways that are fair and just.

Addressing bias in AI-generated code is essential for ensuring that AI tools contribute positively to society and do not inadvertently reinforce harmful stereotypes or inequalities. By focusing on diverse training datasets, algorithmic transparency, bias detection, and ethical AI development practices, stakeholders can mitigate the risks associated with biased AI and promote the development of fair and equitable software solutions.

# 4. Accountability and Responsibility in AI-Generated Code

## 4.1 Error and Liability

As AI tools like GitHub Copilot and ChatGPT become increasingly integrated into the software development process, questions about who is responsible when AI-generated code leads to errors, bugs, or security vulnerabilities become crucial. These issues raise significant challenges in assigning accountability between the developer and the AI tool.

## Who Bears the Responsibility?

- When AI-generated code results in a bug or security flaw, determining who is at fault can be complex. The developer who implemented the AI-generated code is often the first to be held accountable since they made the final decision to include the code in the software. However, if the error arises directly from the AI's suggestion, it might seem unjust to hold the developer solely responsible, especially if they had little knowledge or control over the AI's inner workings.
- AI tool providers may also bear some responsibility, particularly if the AI's output is inherently flawed due to deficiencies in the training data or algorithms. If the AI tool is marketed as a reliable assistant, there could be an expectation that the suggestions it provides are accurate and safe to use. In such cases, liability could extend to the developers of the AI tool itself.

## Legal and Contractual Implications:

- The legal landscape for AI-generated code is still evolving, and there are no clear-cut answers regarding liability. Contracts between developers and organizations might need to specify the extent of the developer's responsibility when using AI tools. For example, clauses could outline the developer's duty to review and test AI-generated code before implementation.
- In the absence of specific legal frameworks, organizations might rely on established software liability principles, where the entity deploying the software (often the developer or the organization) is held liable for any damages caused by the software, regardless of whether the code was written by a human or generated by an AI tool.

## Challenges of Assigning Accountability:

- The challenges of assigning accountability are exacerbated by the "black-box" nature of many AI models. Developers using these tools might not fully understand how the AI generates its suggestions, making it difficult to assess the quality or safety of the code.
- Additionally, the collaborative nature of AI-assisted coding blurs the lines between human and machine input, making it hard to distinguish which parts of the code are the result of human decisions and which are directly generated by the AI.

# 4.2 Ethical Responsibility

Beyond legal and contractual responsibilities, there is an ethical dimension to the use of AI tools in software development. Developers must consider the potential consequences of their reliance on AI-generated code, particularly when it comes to the quality and safety of the software they produce.

## Verification of AI-Generated Code:

- Ethically, developers have a responsibility to verify every AI-generated suggestion before integrating it into their projects. This includes thoroughly reviewing the code for potential errors, security vulnerabilities, and adherence to best practices. Blindly trusting AI suggestions can lead to significant risks, especially in critical applications where software bugs or security flaws could have severe consequences.
- Verification also extends to understanding the AI's limitations. Developers should be aware of the potential biases, inaccuracies, or blind spots that might exist in the AI model. For instance, if an AI tool is known to generate code with certain types of bugs or vulnerabilities, developers must take extra precautions when reviewing those suggestions.

## Consequences of Blind Trust:

- Blindly trusting AI-generated code can lead to a false sense of security, where developers assume that the AI's output is inherently correct. This can result in overlooked errors, leading to software failures, security breaches, or even legal liability if the code causes harm to users or third parties.
- The ethical implications extend to the broader impact on the software development community. If developers consistently rely on AI without critical scrutiny, it could lead to a degradation of coding skills, as developers may become less proficient in identifying and resolving issues themselves. This dependency on AI tools could ultimately reduce the overall quality of software produced by the industry.

### Balancing Efficiency and Responsibility:

- While AI tools offer significant efficiencies in the coding process, developers must balance these benefits with their ethical responsibility to produce reliable, secure, and high-quality software. This means using AI tools as an aid rather than a crutch, maintaining a vigilant and critical approach to the code they produce.
- Organizations can support ethical responsibility by providing training on the use of AI tools, promoting a culture of code review and testing, and encouraging developers to prioritize quality and security over speed.

Accountability and responsibility in AI-generated code involve navigating the complex intersection of legal liability and ethical obligations. Developers and organizations must work together to ensure that AI tools are used in a way that promotes safe, secure, and high-quality software development. This requires a clear understanding of who is responsible when things go wrong, as well as a commitment to ethical practices that prioritize verification, scrutiny, and the careful integration of AI-generated code.

# 5.    Privacy Concerns in AI-Generated Code

## 5.1 Data Usage in AI Training

AI models like GitHub Copilot and ChatGPT are trained on vast amounts of data sourced from the internet, including public code repositories, forums, documentation, and other text sources. This data is often collected on large scale, with little to no input or consent from the original creators. This practice raises significant privacy concerns, particularly when personal information or proprietary code is included in the training data.

### Collection Without Consent:

- The data used to train AI models is typically harvested from publicly accessible sources. However, just because data is public does not mean it is free for all uses. Creators of the data—whether they are software developers, writers, or content creators—may not have anticipated or consented to their work being used to train AI models.
- This lack of consent can lead to privacy violations, especially if the data includes identifiable information, confidential business logic, or proprietary algorithms. For instance, a developer might inadvertently share sensitive information in a public forum or repository, which could then be scraped and used to train an AI model without their knowledge.

### Implications for Privacy:

- The use of data without consent can erode trust in online platforms, as individuals may become reluctant to share information publicly if they fear it could be used in ways they did not intend. This is particularly concerning in industries where privacy and confidentiality are paramount, such as healthcare, finance, or legal services.

- Additionally, there are legal implications related to the unauthorized use of data, particularly in jurisdictions with strict data protection laws like the European Union's General Data Protection Regulation (GDPR). These regulations require explicit consent for the use of personal data, and AI models trained on data without such consent could be in violation of these laws.

## Ethical Considerations:

- Beyond legal concerns, the ethical implications of using data without consent are significant. AI developers and companies must consider whether it is morally acceptable to use someone's work to train AI models without their permission, particularly if the data includes sensitive or personal information.
- Companies developing AI tools have a responsibility to ensure that the data they use is collected and used in an ethical manner. This might involve obtaining explicit consent from data creators, anonymizing data to protect privacy, or being transparent about the data sources used.

## 5.2 AI and Sensitive Data

The interaction between AI-generated code and sensitive or personal data is another area of concern. AI tools can generate code that directly interacts with databases, APIs, or systems that store or process sensitive information, such as personal health data, financial records, or legal documents. This raises several risks related to privacy and data protection.

## Risks of Exposure:

- AI-generated code could inadvertently expose sensitive data if it is not properly examined or if it contains security vulnerabilities. For example, an AI tool might generate code that interacts with a database but fails to adequately secure the connection or validate input, leading to potential data breaches.
- There is also the risk that AI tools could generate code that mishandles sensitive data, such as logging personal information in an insecure manner or failing to anonymize data before it is processed or transmitted. These types of errors could lead to the unauthorized disclosure of personal information, with serious consequences for individuals and organizations.

## Challenges in Managing Sensitive Data:

- Managing sensitive data with AI-generated code is challenging because developers may not fully understand the complexities of data protection and privacy regulations. AI tools might generate code that seems functional but does not comply with legal requirements for handling sensitive information, such as encryption standards, access controls, or data retention policies.
- Additionally, AI-generated code might lack the necessary context to make informed decisions about data handling. For example, an AI tool might generate code that processes sensitive health information without understanding the specific legal or ethical requirements associated with such data.

## Mitigation Strategies:

- To mitigate the risks associated with AI-generated code interacting with sensitive data, developers must take a proactive approach to privacy and security. This includes thoroughly reviewing AI-generated code, conducting security audits, and implementing privacy-enhancing technologies like data anonymization, encryption, and access controls.
- Developers should also be trained to recognize the unique challenges of working with sensitive data and should be aware of the legal and ethical implications of their code. Organizations can support this by providing guidelines and best practices for using AI tools in privacy-sensitive environments.

## Transparency and Accountability:

- AI tool providers should prioritize transparency regarding how their models interact with sensitive data. This includes documenting how the AI was trained, what data sources were used, and how the tool handles sensitive information. Providing developers with clear guidance on the limitations and risks of using AI-generated code in sensitive contexts can help prevent privacy violations.
- Ultimately, developers and organizations must be held accountable for ensuring that AI-generated code complies with privacy regulations and ethical standards. This means taking ownership of the code they implement, even when it is generated by an AI tool, and being prepared to address any privacy concerns that arise.

Privacy concerns in AI-generated code are multifaceted, involving both the data used to train AI models and the interaction of AI-generated code with sensitive or personal information. Addressing these concerns requires a combination of legal, ethical, and technical approaches, including obtaining consent for data use, ensuring the security of sensitive data, and maintaining transparency in AI development. By proactively addressing privacy issues, developers and organizations can harness the power of AI tools while protecting the privacy and rights of individuals.

# 6. Impact on the Job Market and Developer Roles

## 6.1 Automation and Job Displacement

The integration of AI tools like GitHub Copilot and ChatGPT into software development processes has the potential to significantly impact the job market, particularly concerning job displacement and automation of certain roles.

## Potential for Automation:

- **Routine Coding Tasks:** AI tools can automate routine and repetitive coding tasks such as code generation, bug fixing, and code completion. This automation can increase efficiency but may also reduce the need for developers to perform these tasks manually. For instance, AI tools can quickly generate boilerplate code or suggest improvements, which may lead to fewer entry-level positions or roles focused on these routine tasks.
- **Code Review and Testing:** AI can assist in code review by identifying potential issues and vulnerabilities, as well as in automated testing by generating test cases and verifying code functionality. This could impact roles that focus on manual code review and testing, potentially reducing the demand for such positions.

## Roles Most at Risk:

- **Junior Developers:** Entry-level developers who primarily perform routine coding and testing tasks are at higher risk of displacement due to AI automation. As AI tools become more proficient in these areas, the demand for junior developers to handle repetitive tasks may decline.
- **Quality Assurance (QA) Testers:** QA testers who focus on manual testing and bug identification might face reduced job opportunities as AI-driven tools become more adept at automating these processes.
- **Support and Maintenance:** Roles focused on maintaining legacy systems or performing routine updates might also be affected as AI tools streamline these tasks and reduce the need for human intervention.

# 6.2 Shift in Developer Skills

As AI tools become more integrated into the software development lifecycle, the skills required for developers are likely to evolve, shifting from traditional coding skills to more specialized skills related to managing and overseeing AI tools.

## From Coding to Oversight:

- **AI Tool Management:** Developers will need to acquire skills in managing and configuring AI tools, including understanding how these tools generate code and ensuring their outputs align with project requirements and standards.
- **Quality Assurance of AI Outputs:** Instead of focusing solely on writing code, developers will need to develop expertise in evaluating and validating AI-generated code to ensure its accuracy, security, and compliance with best practices. This includes understanding the limitations of AI tools and applying human judgment to correct or refine AI-generated suggestions.

## New Skill Sets:

- **Data Literacy:** Understanding how AI tools are trained and how data influences their outputs will become increasingly important. Developers may need skills in data management, data privacy, and interpreting data-driven insights.
- **Ethics and Governance:** With AI tools introducing new ethical considerations, developers will need to be knowledgeable about the ethical implications of AI and how to address them in their work. This includes understanding bias, privacy concerns, and ensuring responsible AI usage.

## Interdisciplinary Knowledge:

- **Collaboration with AI Specialists:** Developers may work more closely with AI specialists, data scientists, and ethics professionals to ensure that AI tools are used effectively and responsibly. This collaboration requires an understanding of AI principles and the ability to communicate across different areas of expertise.

## 6.3 Long-Term Implications

The long-term impact of AI on the job market and developer roles encompasses both potential benefits and challenges, influencing the broader societal landscape of the software development industry.

### Augmentation of Human Capabilities:

- **Increased Productivity:** AI tools can augment human capabilities by handling routine tasks, allowing developers to focus on more complex and creative aspects of software development. This could lead to increased productivity and the ability to tackle more ambitious projects.
- **Innovation and New Opportunities**: The rise of AI could drive innovation, creating new opportunities for developers to work on advanced projects, such as developing new AI tools, creating applications that leverage AI, or exploring emerging technologies like blockchain or quantum computing.

### Job Losses and Transition:

- **Potential for Job Losses:** While AI can enhance productivity, it may also lead to job losses in certain areas, particularly for roles that are highly automatable. The displacement of jobs could impact developers who specialize in routine tasks, requiring them to adapt to new roles or industries.
- **Reskilling and Upskilling:** To mitigate job losses, there will be a need for reskilling and upskilling programs to help developers transition into new roles or adapt to the changing job market. Educational institutions and employers will play a crucial role in providing training and support for these transitions.

### Societal Impact:

- **Economic Shifts:** The integration of AI into software development may lead to economic shifts, with potential growth in sectors that harness AI and increased competition for roles that require specialized skills. Policymakers and industry leaders will need to address these shifts to ensure equitable opportunities and support for affected workers.
- **Ethical and Social Considerations:** The broader societal impact will also involve ethical and social considerations, such as ensuring fair access to new opportunities and addressing potential biases in AI tools that could affect various groups differently.

The impact of AI on the job market and developer roles is multifaceted, involving both opportunities and challenges. While AI tools can automate routine tasks and enhance productivity, they also bring risks of job displacement and require a shift in developer skills. The long-term implications will depend on how effectively developers, organizations, and policymakers address these changes, balancing the benefits of AI with the need for responsible management and support for affected workers.

# 7.    The Role of Regulation and Governance in AI-Driven Software Development

## 7.1 Need for Regulation

As AI technologies become increasingly integrated into software development, the necessity for robust regulatory frameworks has become more apparent. Currently, the regulatory landscape for AI in software development is often fragmented and insufficiently comprehensive, leading to several key concerns:

### Lack of Comprehensive Regulation:

- Many jurisdictions lack specific regulations that address the unique challenges posed by AI in software development. Existing laws may not adequately cover issues such as data privacy, algorithmic transparency, or accountability for AI-generated outputs. This gap can lead to inconsistencies in how AI technologies are managed and used across different regions and industries.
- The absence of clear regulations can result in ethical and legal uncertainties, making it difficult for developers and organizations to navigate compliance and ensure responsible use of AI tools.

### Ethical Concerns:

- Ethical issues related to AI, such as bias in AI-generated code, data privacy, and intellectual property rights, require more robust legal frameworks. Without appropriate regulation, there is a risk that unethical practices may proliferate, leading to potential harm to individuals or groups and undermining trust in AI technologies.
- The need for regulation is also driven by the growing concern about the social impact of AI, including the potential for job displacement and the ethical treatment of data used to train AI models.

### Calls for Stronger Legal Frameworks:

- Experts and stakeholders advocate for more comprehensive and enforceable legal frameworks that address the specific needs of AI in software development. This includes regulations that ensure transparency, accountability, and fairness in AI systems, as well as mechanisms for addressing grievances and mitigating risks associated with AI technologies.

## 7.2 Governance Models

To address the challenges and ethical concerns associated with AI in software development, various governance models can be considered. Each model offers different approaches to managing AI technologies and ensuring their responsible use:

### Industry Self-Regulation:

- Voluntary Standards and Best Practices: Industry self-regulation involves the establishment of voluntary standards and best practices by industry groups and professional organizations. This model allows for the development of guidelines that can be tailored to specific sectors and use cases, fostering ethical behavior and innovation while providing flexibility for organizations.

- Ethical Codes and Certification: Self-regulation can also include the creation of ethical codes of conduct and certification programs that developers and organizations can adhere to. These measures can promote responsible AI development and use, although they rely on voluntary compliance and may lack enforcement mechanisms.

## Government Oversight:

- **Legislative and Regulatory Frameworks**: Government oversight involves the creation and enforcement of legislative and regulatory frameworks that address key issues related to AI in software development. This can include laws governing data privacy, algorithmic accountability, and transparency, as well as mechanisms for oversight and enforcement.
- **Regulatory Bodies and Agencies:** Establishing specialized regulatory bodies or agencies can provide dedicated oversight for AI technologies. These entities can develop and enforce regulations, conduct audits, and provide guidance on best practices, ensuring that AI tools are used responsibly and ethically.

## International Cooperation:

- **Global Standards and Agreements**: Given the global nature of AI development and deployment, international cooperation is essential for creating consistent and effective regulations. Global standards and agreements can help harmonize regulations across borders, facilitating collaboration and ensuring that AI technologies meet common ethical and legal standards.
- **Collaborative Platforms and Forums:** International forums and collaborative platforms can bring together stakeholders from different countries to discuss and address global challenges related to AI. These platforms can facilitate knowledge sharing, promote best practices, and coordinate efforts to develop and implement international regulations.

The role of regulation and governance in AI-driven software development is crucial for addressing ethical concerns and ensuring the responsible use of AI technologies. While industry self-regulation and government oversight offer valuable approaches, international cooperation is essential for creating a cohesive and effective regulatory framework. By adopting a combination of these governance models, stakeholders can work towards a balanced and responsible approach to managing AI in software development, fostering innovation while safeguarding ethical standards and public trust.

# 8.  Real-World Examples

## Ethical Challenges in AI - Misinformation and Privacy Concerns in Google's Bard AI and Meta's BlenderBot 3 (2023)

Google's Bard AI, a sophisticated conversational model launched to rival other advanced language models like ChatGPT, and Meta's BlenderBot 3, an innovative chatbot aimed at enhancing social interactions on digital platforms, both encountered significant ethical challenges shortly after their releases. Bard AI quickly drew criticism for generating misleading or factually incorrect information, raising serious concerns about the propagation of misinformation. This issue, rooted in biases embedded within its training data, highlighted the complexities of ensuring AI systems deliver accurate and reliable responses. The potential for widespread misinformation not only posed a threat to public trust in AI technologies but also underscored the critical need for increased transparency and accountability in how AI outputs are produced and verified.

Simultaneously, Meta's BlenderBot 3 grappled with privacy issues when users discovered it could share sensitive personal information during interactions. This raised alarms about data privacy and the ethical handling of user data, as the chatbot's ability to recall and reference previous conversations without explicit consent pointed to significant risks in data management practices. The incident underscored the dangers of AI systems accessing and utilizing personal data in ways that could lead to breaches of user privacy, thereby eroding trust in AI-driven platforms.

Both Bard AI and BlenderBot 3 serve as case studies of the broader ethical implications that arise when deploying AI in content generation and data management. They emphasize the importance of rigorous data quality and bias mitigation strategies to prevent the dissemination of misinformation. Additionally, these cases highlight the necessity of robust verification mechanisms, enhanced transparency in AI operations, and stringent data privacy protocols. Addressing these issues is essential to developing AI systems that are not only technically advanced but also ethically sound, ensuring that AI technologies are deployed in a manner that upholds public trust and safeguards user rights.

# 9. Future Directions in AI Ethics for Software Development

## 9.1 Evolving Ethical Standards

As AI technology continues to advance, the ethical standards governing its use in software development are likely to evolve in response to emerging challenges and opportunities. Key areas where ethical standards might evolve include:

### Dynamic Regulation:

- **Adaptive Frameworks:** Ethical standards for AI may shift towards more adaptive and dynamic regulatory frameworks that can quickly respond to technological advancements and emerging ethical issues. These frameworks might include flexible guidelines that evolve alongside AI capabilities and applications.
- **Global Harmonization:** There may be an increasing push for global harmonization of ethical standards to ensure consistent and fair practices across different regions and industries. International collaborations and agreements could play a crucial role in establishing universal principles for AI ethics.

### Enhanced Transparency and Explainability:

- **AI Explainability:** As AI systems become more complex, there will likely be a greater emphasis on transparency and explainability. Future ethical standards may require more detailed explanations of AI decision-making processes and the ability to audit and understand how AI systems arrive at their conclusions.

- **User Empowerment:** Ethical guidelines may increasingly focus on empowering users with better understanding and control over AI interactions. This could include clearer disclosures about data usage and more accessible mechanisms for users to provide feedback and address concerns.

## Bias and Fairness:

- **Proactive Bias Management:** Evolving standards will likely place greater emphasis on proactive measures for detecting and mitigating biases in AI systems. This could involve more sophisticated techniques for bias auditing and the implementation of continuous monitoring to address emerging biases.
- **Inclusive Design:** Future ethical standards may advocate for more inclusive design practices that consider the diverse needs and perspectives of different user groups. This approach aims to ensure that AI systems serve all users fairly and equitably.

## Ethical AI Development Practices:

- **Ethical Design Principles:** There may be a shift towards integrating ethical design principles into the AI development lifecycle. This could involve incorporating ethical considerations from the initial stages of design and development, rather than addressing them as an afterthought.
- **Accountability and Responsibility:** Future standards may place a stronger emphasis on accountability and responsibility, holding developers and organizations accountable for the ethical implications of their AI systems and ensuring that ethical practices are embedded throughout the development process.

## 9.2 AI Ethics Research

As AI technology continues to evolve, several areas of research will be crucial for advancing our understanding of AI ethics, particularly in the context of software development:

## Algorithmic Fairness and Bias:

- **Bias Detection and Mitigation**: Research into more effective methods for detecting and mitigating bias in AI systems is essential. This includes developing advanced algorithms and techniques for identifying and addressing biases in training data, model outputs, and decision-making processes.
- **Fairness Metrics:** Investigating new metrics and methodologies for measuring fairness in AI systems can help ensure that ethical standards are met. Research in this area should focus on developing comprehensive and actionable fairness metrics that can be applied across various applications.

## AI Explainability and Transparency:

- **Explainable AI (XAI):** Continued research into explainable AI aims to improve our understanding of how AI systems make decisions and provide clearer explanations to users. This includes developing methods for generating interpretable and transparent AI models that can be audited and understood by non-experts.

- **Transparency Tools:** Exploring new tools and techniques for enhancing transparency in AI systems, such as visualization tools and user-friendly interfaces, can help users better understand AI behavior and decision-making processes.

## Ethical Implications of AI in High-Stakes Domains:

- **Sensitive Applications:** Research into the ethical implications of AI in high-stakes domains, such as criminal justice, healthcare, and finance, is critical for addressing potential risks and ensuring responsible use. This includes studying the impact of AI on vulnerable populations and developing guidelines for ethical AI use in these areas.
- **Regulatory and Governance Models**: Investigating effective regulatory and governance models for managing AI technologies, including industry self-regulation and international cooperation, can help establish best practices and standards for ethical AI development.

## Human-AI Interaction and User Trust:

- **User Interaction Studies:** Research into how users interact with AI systems and the factors that influence trust and acceptance can provide insights into improving user experience and addressing ethical concerns. This includes studying user perceptions of AI transparency, accountability, and fairness.
- **Ethical Design Practices:** Exploring ethical design practices and principles that prioritize user well-being and responsible AI development can help guide the creation of AI systems that align with ethical standards and societal values.

The future directions in AI ethics for software development will involve evolving ethical standards to address emerging challenges, such as ensuring transparency, managing bias, and embedding ethical practices into the AI development process. Continued research in areas such as algorithmic fairness, AI explainability, and the ethical implications of AI in high-stakes domains will be crucial for advancing our understanding and ensuring that AI technologies are developed and used responsibly. By addressing these research areas, we can work towards creating AI systems that are ethical, fair, and aligned with societal values.

# 10. Conclusion

This paper explores the ethical issues associated with AI in software development, including questions of authorship and intellectual property, where the ownership of AI-generated code remains a significant concern. It highlights the need to distinguish between human contribution and AI generation, emphasizing the importance of understanding how AI augments human creativity. The paper also addresses bias in AI-generated code, particularly the risk of perpetuating societal biases in sensitive applications, underscoring the necessity of effective bias detection and mitigation. Accountability and responsibility are examined, especially regarding who is liable when AI-generated code results in bugs or security vulnerabilities. Privacy concerns are raised, focusing on the importance of informed consent and robust data protection when training AI models with sensitive data. Additionally, the paper considers the impact of AI tools on the job market and developer roles, noting the shift towards overseeing and managing AI tools rather than performing routine tasks. The need for regulation and governance is also discussed, advocating for robust frameworks to address ethical concerns. Finally, case studies of Google's Bard AI and Meta's BlenderBot 3 provide real-world examples of ethical challenges, offering lessons on improving AI systems and addressing dilemmas.

The integration of AI into software development offers significant opportunities but also presents ethical challenges that must be addressed proactively. As AI technology advances, it is crucial for the software development community to implement best practices for data handling, ensure transparency and accountability, and continuously monitor AI systems to prevent biases and misinformation. Achieving responsible AI integration requires collaboration among developers, researchers, policymakers, and industry leaders. By fostering a culture of transparency and responsibility, embracing ethical guidelines, and promoting ongoing dialogue, research, and policy-making, the community can harness the benefits of AI while mitigating potential risks. Through these collective efforts, we can ensure that AI technologies contribute positively to society, aligning with societal values and serving the greater good.